\documentstyle[preprint,prl,aps,psfig]{revtex}
\tightenlines
\begin{document}
\draft
\title{Magnetic dipole and electric quadrupole responses of 
elliptic quantum 
dots in magnetic fields}
\author{Enrico Lipparini}
\address{Dipartimento di Fisica, Universit\`a di Trento,
and INFM sezione di Trento, 
I-38050 Povo, Italy}
\author{Lloren\c{c} Serra, Antonio Puente}
\address{Departament de F\'{\i}sica, Universitat de les Illes Balears,
E-07071 Palma de Mallorca, Spain}
\date{April 9, 2001}
\maketitle
\begin{abstract}
The magnetic dipole (M1) and electric quadupole (E2) responses of 
two-dimensional quantum dots with an elliptic shape are theoretically 
investigated as a function 
of the dot deformation and applied static magnetic field.
Neglecting the electron-electron interaction we obtain analytical 
results which indicate the existence of four characteristic modes, with 
different $B$-dispersion of their energies and associated strengths.
Interaction effects are numerically studied within the time-dependent
local-spin-density theory, assessing the validity of the non-interacting 
picture.   
\end{abstract}
\pacs{PACS 73.20.Dx, 72.15.Rn}
\narrowtext

\section{Introduction}
The electric dipole (E1) response of quantum dots has deserved much 
attention in recent years, mainly motivated by the measurements of 
far-infrared absorption in these systems \cite{Sik89,Dem90}.
The manifestation of the so called {\em magnetoplasmons} in a
perpendicularly applied magnetic field constitutes an example of collective 
oscillation in finite Fermi systems, physically analogous to those 
existing in metal clusters, atoms or nuclei. In parabolically confined dots 
magnetoplasmons are understood as rigid motions of the electronic center 
of mass, a result that stems from the generalized Kohn's
theorem \cite{Brey89,Mak90,Broi90,Gud91,Ser99}. 
However, 
deviations from parabolicity, such as angular deformations or non-quadratic radial 
behaviour, may result in more complicated absorption patterns which are 
currently being much 
investigated \cite{Pue99,Ull00,Gudxx}.

The understanding of other excitation modes of quantum dots are also 
essential for a proper characterization of these systems. Techniques 
based on resonant inelastic light scattering have already proved extremely 
useful to this purpose \cite{Str94,Loc96,Sch96,Sch98}. In fact,
using polarization selection rules they permit
to disentangle charge-density from spin-density
and single-particle excitations, as well as to discern different 
multipolarity peaks in each channel. The momentum transfer and magnetic 
field dependence of the different excitations of a single sample 
have been recently reported using this technique \cite{Sch98}.
It is worth to mention that 
theoretical analysis of the Raman spectra in circularly symmetric
quantum dots have been presented in 
Refs.\ \cite{Stei99,Stef99,Bar00}.

For elliptical quantum dots 
orbital-current modes and their relationship with the quadrupole response 
were analysed by us in Ref.\ \cite{orbital}.
A characteristic low energy mode, depending on deformation and with a 
conspicuous signal in the magnetic dipole (M1) channel 
was found in Ref.\ \cite{orbital}, while the 
relevance of orbital excitations
in the build-up of the electronic moment of inertia was discussed 
in Ref.\ \cite{inert}. In this paper we focus on the magnetic field 
dependence of the orbital and quadrupole modes, a necessary step to 
understand the M1 and E2 responses in deformed dots and towards the 
description of Raman scattering in symmetry unrestricted 
nanostructures. It will be shown that
four excitation modes with $B$-dependent energies and 
strengths characterize both M1 and E2 reponses in elliptic dots. This 
result follows from analytical calculations in the so-called 
non-interacting deformed-harmonic-oscillator (NIDHO) model, as well as 
for interacting electrons in the local-spin-density approximation 
(LSDA). Therefore, it constitutes a robust picture against interactions
in a somewhat similar way as Kohn's modes are.

This paper is organized as follows: Sec.\ II is devoted to the 
NIDHO model for M1 and E2 responses; in Sec.\ III we present the
time-dependent LSDA results for the same channels; finally, Sec.\ IV
presents the conclusions.    

\section{The NIDHO model}

\subsection{The analytic solution of the ground state}
We consider a system of non-interacting electrons, confined
to the $xy$ plane by an anisotropic parabola \cite{units}
\begin{equation}
v^{\em (conf)}(x,y)=\frac{1}{2}(\omega_x^2 x^2 + \omega_y^2 y^2)\; .
\label{eq1}
\end{equation}
The mean parabola coefficient $\omega_0=(\omega_x+\omega_y)/2$ is 
parametrized as usual, in terms of the Wigner-Seitz radius $r_s$
and electron number $N$ as $\omega_0^2=1/r_s^3\sqrt{N}$. In the
following we shall label 
the dot deformation with the parameter 
$\delta=\omega_y/\omega_x$ (ranging from 0 to 1). 
Taking into account a uniform magnetic field ${\bf B}=B {\bf e}_z$, 
the system Hamiltonian reads 
\begin{equation}
H = \sum_{i=1}^N{\left[
\frac{1}{2}\left( {\bf p_i}+\gamma {\bf A}({\bf r}_i) \right)^2 
+ v^{\em (conf)}({\bf r}_i)\right]}
+g^* \gamma B S_z\; , 
\label{eq2}
\end{equation}
where $\gamma=e/c$ (we assume Gaussian
magnetic fields) and within the symmetric gauge 
${\bf A}({\bf r})=B/2(-y,x)$. The last piece is the Zeeman term,
depending on the total spin $S_z$ 
and the effective gyromagnetic factor $g^*$ \cite{note}. Obviously
this is a one-electron picture, in which the relevant one-electron
Hamiltonian is $h=h_{xy}+g^* \gamma B s_z$. 
Because of the magnetic field, the spatial part $h_{xy}$ deviates from 
a simple harmonic oscillator problem. Namely
\begin{eqnarray}
h_{xy} &=&\frac{1}{2}(p_x^2+p_y^2)+
\frac{\omega_c}{2}(xp_y-yp_x)\nonumber\\
&+& \frac{1}{2}(\tilde{\omega}_x^2 x^2+\tilde{\omega}_y^2 y^2)\; , 
\label{eq3}
\end{eqnarray}
with $\tilde{\omega}_{x}^2=\omega_{x}^2+\frac{1}{4}\omega_c^2$ and
$\tilde{\omega}_{y}^2=\omega_{y}^2+\frac{1}{4}\omega_c^2$
the new parabola coefficients, given in terms of the cyclotron
frequency $\omega_c=eB/c$.

The nontrivial problem posed by Eq.\ (\ref{eq3}) has been elegantly
solved in an analytical way, by Dippel {\em et al.} \cite{Dip94}  
and Madhav and Chakraborty \cite{Mad94}
in the context of atomic physics and quantum dots, respectively. 
In the following, we will refer to the derivation by 
Dippel {\em et al.}.  They introduce a similarity 
transformation to a new Hamiltonian $h_3=U^{-1}hU$, where
$U=e^{i\alpha xy}e^{i\beta p_xp_y}$.  The parameters $\alpha$ and 
$\beta$ are chosen in order to obtain two decoupled oscillators 
in the $x$ and $y$ directions
\begin{equation}
h_3=\frac{p_x^2}{2M_1}+\frac{p_y^2}{2M_2}+
\frac{M_1}{2}\omega_1^2x^2+\frac{M_2}{2}\omega_2^2y^2\; .
\label{eq4}
\end{equation}
The reader is addressed to Ref.\ \cite{Dip94} for the detailed 
expressions of $\alpha$, $\beta$, $M_1$, $M_2$, $\omega_1$ and 
$\omega_2$ in terms of the original Hamiltonian parameters.

Most importantly, we can now express the eigenstates of $h_{xy}$
in terms of those of $h_3$. The latter are simply products
of one-dimensional oscillator functions with a given 
number of quanta, i.e.,
\begin{eqnarray}
h_3 \Phi_{n_1n_2}(x,y) &=& E_{n_1n_2} \Phi_{n_1n_2}(x,y) \nonumber\\
\Phi_{n_1n_2}(x,y) &=& \phi_{n_1}(x)\phi_{n_2}(y)\; ,
\label{eq5}
\end{eqnarray}
while the eigenstates of $h_{xy}$ are given by 
\begin{eqnarray}
h_{xy} \Psi_{n_1n_2}(x,y) &=& E_{n_1n_2} \Psi_{n_1n_2}(x,y)\; ,\nonumber\\
\Psi_{n_1n_2}(x,y) &=& U\, \Phi_{n_1n_2}(x,y)\; .
\label{eq7}
\end{eqnarray}
Notice that the energy eigenvalues of $h_{xy}$ coincide with those
of $h_3$ and that they are
\begin{equation}
E_{n_1n_2} = (n_1+\frac{1}{2})\omega_1 + (n_2+\frac{1}{2})\omega_2\; .
\label{eq8}
\end{equation}
Having the one-electron states it is then a simple matter to obtain the 
$N$ electron ground state of the NIDHO model just by filling the $N$ 
lower energy orbitals.

\subsection{M1 response}

The excitations within the NIDHO model will correspond to independent 
particle-hole (ph) transitions in the level scheme of the ground state. 
Taking the orbital 
angular momentum $\ell_z$ as excitation operator the M1 strength function
will be given by
\begin{eqnarray}
S_{\rm M1}(\omega) &=& \sum_{hp}{
f_h (1-f_p)\,
\left\vert \langle n_1^h n_2^h | U^{-1}\ell_z U | n_1^p n_2^p \rangle 
\right\vert^2}\times
\nonumber\\
&& \qquad \delta(E_p-E_h-\omega) \; ,
\label{eq9}
\end{eqnarray}
where we have denoted the electron states $\Phi_{n_1n_2}$ as 
$| n_1 n_2 \rangle$ and the occupation numbers 
are given by the $f$ factors.

To evaluate the matrix element in Eq.\ (\ref{eq9}) we first need to 
obtain the transformed operator. This can be accomplished by using the 
Baker-Haussdorf lemma for unitary transformations \cite{Sakxx}, which
introduces a nested conmutators expansion. A straightforward calculation
yields
\begin{equation}
U^{-1}\ell_zU=(1-2\alpha\beta)\ell_z+\beta(1-\alpha\beta)(p_x^2-p_y^2)
+\alpha(x^2-y^2)  \; .
\label{eq10}
\end{equation}
Position and momentum operators may now be transformed to creation and 
anihilation ones
\begin{eqnarray}
x &=& \left(\frac{1}{2M_1\omega_1}\right)^\frac{1}{2} (a^+_1+a_1) \nonumber\\ 
p_x &=& \left(\frac{M_1\omega_1}{2}\right)^\frac{1}{2}(a^+_1-a_1)  \; , 
\end{eqnarray}
with similar relations for $y$ and $p_y$ in terms of $a_2$ and $a_2^+$. 
The matrix elements are therefore
written in terms of the creation and anihilation operator ones
\begin{eqnarray}
\langle n_\alpha | a^+ | n_\beta \rangle &=& \sqrt{n_\beta+1} \,
\delta_{n_\alpha n_\beta+1}\nonumber\\
\langle n_\alpha | a | n_\beta \rangle &=& \sqrt{n_\beta} \,
\delta_{n_\alpha n_\beta-1} \; .
\label{eq11}
\end{eqnarray}

From the calculation it emerges that there are only four allowed
transition energies: $\omega_{11}=2\omega_1$, 
$\omega_{22}=2\omega_2$, $\omega_+=\omega_1+\omega_2$ and 
$\omega_-=|\omega_1-\omega_2|$ whose M1 strengths are
\begin{eqnarray}
S_{\rm M1}(\omega_{11}) &=&
{(\alpha-\beta(1-\alpha\beta)M_1^2\omega_1^2)^2
\over 4M_1^2 \omega_1^2 }\sum_{\rm occ.\ levels}{(1-f_{n_1+2\, n_2})(n_1+1)(n_1+2)} \nonumber\\
S_{\rm M1}(\omega_{22}) &=&
{(\alpha-\beta(1-\alpha\beta)M_2^2\omega_2^2)^2
\over 4M_2^2 \omega_2^2 }\sum_{\rm occ.\ levels}{(1-f_{n_1\, n_2+2})(n_2+1)(n_2+2)} \nonumber\\
S_{\rm M1}(\omega_+) &=&{(1-2\alpha\beta)^2\over 4M_1 M_2 \omega_1\omega_2}
(M_2\omega_2- M_1\omega_1)^2 
\times
\nonumber\\
&&\!\!\!\!\!\!\!\!\!\!\!\!\!\!\!
\sum_{\rm occ.\ levels}{(1-f_{n_1+1\, n_2+1})(n_1+1)(n_2+1)}
\nonumber\\
S_{\rm M1}(\omega_-) &=&{(1-2\alpha\beta)^2\over 4M_1 M_2 \omega_1\omega_2}
(M_2\omega_2+ M_1\omega_1)^2 
\times
\nonumber\\
&&\!\!\!\!\!\!\!\!\!\!\!\!\!\!\!
\sum_{\rm occ.\ levels}{(1-f_{n_1+1\,n_2-1})(n_1+1)n_2}
\; .
\label{eq12}
\end{eqnarray}
Result (\ref{eq12}) provides a clear interpretation of the 
different modes:
$\omega_{11}$ corresponds to
the absorption of two $x$-oscillator quanta, $\omega_{22}$ absorbs two 
$y$ quanta while $\omega_+$ takes one quantum in each 
oscillator. By contrast, $\omega_-$ is associated with the absorption 
of one $x$ quantum accompanied with the emission of a $y$ quantum. 
We remind the 
reader that our convention ($\omega_y\leq\omega_x$) renders the 
absorption of a $y$ quantum accompanied by the emission of an $x$ quantum 
energetically forbidden.

Figure 1 depicts the absorption energies and associated intensities
$\omega S(\omega)$ as a function of the magnetic 
field for different deformations in a quantum dot with $N=6$ electrons 
and $r_s=1.51$. For $\delta\geq 0.7$, when the system is just slightly
deformed ($\delta=1$ corresponds to the circular case), 
only the $\omega_+$ mode is active at low $B$'s. As the magnetic field is 
increased, very rapidly the
$\omega_{11}$ and $\omega_{22}$ modes gain strength and disperse in 
energy. Eventually, for very large magnetic fields  
$\omega_{22}$ becomes the lowest energy mode and carries most of 
the strength. 
 
The $\omega_-$ mode deserves a special discussion.
Deformed dots at low deformation and magnetic field are characterized 
by the same oscillator occupancy numbers of the corresponding circular 
limit. In the case of magic number dots 
($N={\cal N}({\cal N}+1)$ with ${\cal N}$ an integer number specifying 
the last occuppied shell) the 
$\Delta n_1=1$, $\Delta n_2=-1$ ph transitions are Pauli-blocked
thus completely inhibiting the $\omega_-$ mode. 
This excitation is switched on 
(with a sudden increase in strength) only when the deformation or
$B$ are high enough to break the closed shell structure, which 
takes place
in first instance when
\begin{equation}
\frac{\omega_2}{\omega_1} =
\frac{\cal N}{{\cal N}+1}\; .
\end{equation}
Figure 2 displays the regions in the $(\delta,B)$ plane in which 
the $\omega_-$ mode is active, for different dot sizes.
This effect is absent
for non-magic electron numbers since there is no Pauli blocking for 
these systems and, as a consequence, $\omega_-$ is active at all 
$\delta$ and $B$.
For $N=12$, 20, \dots, subsequent changes in the ground state level
structure give also rise to enhancements of the $\omega_-$ strength
by allowing an increasing number of ph transitions. 
The mechanism discussed here is also responsible for strong 
enhancements of the moment of inertia in these systems \cite{inert}.

The general situation in Fig.\ 1 is qualitatively very similar for the 
deformations $\delta=0.9$ and $0.7$, we remark however that for 
$\delta=0.7$ the {\em splitting} $\omega_{11}-\omega_{22}$ is higher at low $B$ 
and the $\omega_-$ mode switches on at a smaller magnetic field. 
For large deformations ($\delta=0.4$) the $\omega_-$ mode is active even at 
$B=0$. Notice also that the splitting $\omega_{11}-\omega_{22}$ is so high 
that $\omega_{22}$ approaches $\omega_-$ at vanishing magnetic fields.
A common feature to all deformations is that  $\omega_{22}$
dominates the response at high $B$'s. Altogether, the NIDHO model provides a quite 
interesting scenario of the M1 channel, with the emergence of novel
modes and big strength transfers between them as a function of the magnetic 
field and deformation. 

\subsection{E2 response}

The E2 strength is given in terms of the quadrupole 
operator $xy$ by
\begin{eqnarray}
S_{\rm E2}(\omega) &=& \sum_{hp}{
f_h (1-f_p)\,
\left\vert \langle n_1^h n_2^h | U^{-1} xy U | n_1^p n_2^p \rangle 
\right\vert^2}\times
\nonumber\\
&& \qquad \delta(E_p-E_h-\omega) \; .
\label{eq14}
\end{eqnarray}
In a way similar to the M1 analysis we may obtain the E2 matrix 
elements. After a straightforward calculation one finds the 
same four modes of the M1 channel manisfesting in the E2 spectra,
although with different strengths (see third column in Fig\ 1). 
The detailed expressions are
\begin{eqnarray}
S_{\rm E2}(\omega_{11}) &=&
\frac{\beta^2}{4} \sum_{\rm occ.\ levels}{(1-f_{n_1+2\, n_2})(n_1+1)(n_1+2)}
\nonumber\\
S_{\rm E2}(\omega_{22}) &=&
\frac{\beta^2}{4} \sum_{\rm occ.\ levels}{(1-f_{n_1\, n_2+2})(n_2+1)(n_2+2)}
\nonumber\\
S_{\rm E2}(\omega_+) &=&
{(1-\beta^2 M_1 M_2 \omega_1 \omega_2 )^2\over M_1 M_2 \omega_1\omega_2}\times
\nonumber\\
&&\!\!\!\!\!\!\!\!\!\!\!\!\!\!\!
\sum_{\rm occ.\ levels}{(1-f_{n_1+1\, n_2+1})(n_1+1)(n_2+1)}
\nonumber\\
S_{\rm E2}(\omega_-) &=&
{(1+\beta^2 M_1 M_2 \omega_1 \omega_2 )^2\over M_1 M_2 \omega_1\omega_2}\times
\nonumber\\
&&\!\!\!\!\!\!\!\!\!\!\!\!\!\!\!
\sum_{\rm occ.\ levels}{(1-f_{n_1+1\,n_2-1})(n_1+1)n_2}
\; .
\label{eq15}
\end{eqnarray}

We notice that the 
share of strength as a function of the magnetic field and deformation 
is similar to the magnetic dipole case. Several differences may, however, 
be remarked. Firstly, the  
$\omega_+$ mode is relatively enhanced for all $\delta$ and $B$
as compared to the results for the M1 channel. For instance
the intensity ratio $I_{\rm E2}(\omega_+)/I_{\rm E2}(\omega_-)$ 
is enhanced by a factor $\approx 4$ 
at $\delta=0.4$ and $B=0$.
It is also worth to point out that the 
$\omega_{11}$ and $\omega_{22}$ intensities are reverted with respect to 
the M1 result. Finally, at 
large magnetic fields the low energy $\omega_{22}$ mode 
is no longer the dominant one since both $\omega_{11}$ and $\omega_-$
show a higher intensity.

As remarked above
the number of electrons in Fig.\ 1 ($N=6$) corresponds at low 
deformations $\delta\to 1$ to a closed shell quantum dot.  
When analysing
the M1 and E2 responses of dots that correspond to open shells in the circular
limit the results are quite similar to those in this Figure, but 
with the important difference 
that the $\omega_-$ mode is already active at $B=0$ since
Pauli blocking is not effective in this case.

\section{The LSDA approach}

\subsection{The method}

To analyze the role of electron-electron interactions in a microscopic 
formalism we resort to the 
LSDA version of density functional theory. This theory provides a practical
way to include electronic exchange and correlation effects by relying on 
exact calculations for the uniform electron gas. At different levels, 
density functional theory has been applied by many authors to describe 
the ground state \cite{Fer94,Hei95,Kos98,Pi98,Hi99}
and excitations of quantum dots \cite{Ser99,Pue99,Ull00,Lip99}. 
The most refined version
is the so-called current-density functional, first applied to 
quantum dots by Ferconi and Vignale \cite{Fer94}, that was recently 
used to describe the edge reconstruction in these systems
for large magnetic fields \cite{Reixx}. Nevertheless, 
current terms are known to be rather 
small for moderate magnetic fields and will be neglected in this work
in which we shall resort to the approach of Ref.\ \cite{Pue99}, developed
for the treatment of noncircular nanostructures.

The description of excitations in deformed nanostructures constitutes a
highly nontrivial task, mainly because of the lack of symmetry not allowing 
the analytic integration of angular variables as in circular systems.
For this reason we shall use the time-dependent LSDA to obtain the time
evolution, following an initial perturbation, of relevant expectation 
values.
The set of single-particle orbitals $\{ \varphi_i({\bf r})\}$ evolves in 
time as
\begin{equation}
i{\partial\over\partial t} \varphi_{i\eta}({\bf r},t) =
h_\eta[\rho,m]\, \varphi_{i\eta}({\bf r},t)\;,  
\label{eqh}
\end{equation}
where $\eta=\uparrow,\downarrow$ is the spin  index, and total density 
and magnetization are given in terms of the spin densities 
$\rho_\eta({\bf r})=\sum_i{|\varphi_{i\eta}({\bf r})|^2}$, by 
$\rho=\rho_\uparrow+\rho_\downarrow$ and 
$m=\rho_\uparrow-\rho_\downarrow$, respectively. The Hamiltonian $h_\eta$ 
in equation (\ref{eqh}) contains, besides of kinetic energy, 
the confining 
$v^{\em (conf)}({\bf r})$, Hartree 
$v^{(H)}({\bf r})=\int{d{\bf r}' \rho({\bf r}')/{|{\bf r}-{\bf r}'|}}$
and exchange-correlation 
$v^{(xc)}_\eta({\bf r})=
{\partial\over\partial\rho_\eta}{\cal E}_{xc}(\rho,m)$ potentials.
The exchange-correlation energy density ${\cal E}_{xc}(\rho,m)$ has been 
described as in Refs.\ \cite{Ser99,Pue99,Ull00,Lip99}. This
technique was already
used by us in Ref.\ \cite{orbital} to describe orbital ($\omega_-$) 
excitations in quantum dots at zero magnetic field.

An initial perturbation of the orbitals 
$\varphi'({\bf r})={\cal P}\varphi({\bf r})$ models the interaction with the 
external field. The unitary operator ${\cal P}$ is given in terms of a displacement
field ${\bf u}({\bf r})$ as 
\begin{equation}
{\cal P} = \exp{\left[
i {{\bf u}({\bf r})\cdot{\bf p} } \right] }\; .
\label{eq16}
\end{equation}
The appropriate fields for the M1 and E2 channels are
${\bf u}_{\rm M1}=\lambda\,r\, {\bf e}_\theta$ and 
${\bf u}_{\rm E2}=\lambda \nabla(xy)$, where $\lambda$ is a small
parameter that guarantees the linear regime.
Actually, these operators correspond to a rigid rotation (M1)
and a quadrupole distortion (E2) of the electronic orbitals. 
After the initial distortion we keep track of 
$\langle\sum_i\ell^{(i)}_z\rangle(t)$ or $\langle\sum_i x_iy_i\rangle(t)$
and a subsequent frequency analysis of the signal provides the 
absorption energies and their associated strengths \cite{Pue99}.

\subsection{LSDA results and discussion}

Figure 3 summarizes the M1 and E2 spectra of the $N=6$ electron dot within
time-dependent LSDA.  Quite remarkably many features of the NIDHO model are 
also present in the LSDA spectra. In fact we can identify the four modes, 
labeled in Fig.\ 3 by analogy with the NIDHO results, although in 
general they lie at different energies because of the interactions.

Focusing on the M1 channel, we notice for $\delta=0.9$ the emergence
of the low energy mode $\omega_-$ and its crossing with the $\omega_{22}$
mode. This crossing takes place at $B\approx 1.5$~T for both 
$\delta=0.9$ and $\delta=0.7$ and at $B\approx 1$~T for  
$\delta=0.4$.
The dominance of $\omega_{22}$ at 
large $B$'s is also in very good agreement with the NIDHO prediction. 
Quite remarkably, for $\delta=0.7$ the $\omega_-$ mode is switched on
even at $B=0$ as a result of the interactions. At $\delta=0.4$ the closeness
of $\omega_{22}$ and $\omega_-$ at low magnetic fields predicted in 
the non-interacting case is nicely seen
in Fig.\ 3. It is also worth to mention the strong quenching of the 
$\omega_{11}$ mode within LSDA.

The E2 results also reflect to a great extent the systematics 
predicted by the NIDHO model. In general the high energy modes 
$\omega_+$ and $\omega_{11}$ are more important in the E2 spectra than 
in the M1 ones, especially at large magnetic fields. In some cases the 
LSDA strengths show some small fragmentations that we attribute to 
ph effects (Landau damping). For instance, this is quite clear in the 
$\omega_{11}$ mode at 3 and 4 teslas.

\section{Summary and conclusions}

The M1 and E2 channels of elliptic quantum dots are characterized by 
four modes whose energies and strengths vary with magnetic field and
deformation. Their main characteristics are already seen in a 
non-interacting model, although interactions shift the mode energies 
and introduce several minor differences. The main trends seen in 
Figs.\ 1-3 may be summarized as follows:
\begin{enumerate}

\item[a)] Three modes $\omega_{11}$, $\omega_+$ and $\omega_-$ 
have a positive dispersion relation with $B$ while the other
$\omega_{22}$ exhibits a negative one. 
At low magnetic fields and deformations $\omega_{11}$, $\omega_+$ 
and $\omega_{22}$ are very close, 
while $\omega_-$ lies at a much lower energy. As the deformation is 
increased the three upper modes separate in energy (at $B \approx 0$), 
the lower of them $\omega_{22}$ coming close to $\omega_-$.

\item[b)] At low $B$'s
the M1 strength of closed shell dots 
is exhausted by $\omega_+$ while, when increasing the deformation 
$\omega_-$ is also active. This mode is also switched on when increasing 
the magnetic field
for a fixed deformation as a consequence of the Pauli blocking mechanism.
Generally, a crossing between $\omega_-$ and 
$\omega_{22}$ occurs for intermediate magnetic fields.
At large magnetic fields the low energy $\omega_{22}$ mode eventually 
takes all the strength.   
 
\item[c)] The E2 spectra are similar to the M1 ones, although an 
important amount of 
strength shifts to the high energy modes $\omega_+$, $\omega_-$ 
and $\omega_{11}$ as compared to the magnetic dipole channel. 
Contrary to the M1 case
the E2 absorption does not show a clear dominant peak at high $B$'s.

\end{enumerate}

\acknowledgments
This work was supported in part by Grant No.\ PB98-0124 from DGESeIC, 
Spain.


\begin{figure}[f]
\caption{Results within the NIDHO model for a $N=6$ electron dot with 
$r_s=1.51$.
Left panels display the mode energies in H$^*$ while central and right ones show 
the M1 and E2 absorption cross sections $\omega S(\omega)$ as a function 
of the magnetic field. Each row corresponds to a different dot 
deformation.}
\end{figure}

\begin{figure}[f]
\caption{The lines show the boundary between regions for
$\omega_-$ mode in the NIDHO model and different dot sizes. 
Left and right of each curve correspond to 
$\omega_-$ being active or non-active (Pauli blocked) respectively.
}
\end{figure}

\begin{figure}[f]
\caption{M1 an E2 absorption intensities within LSDA for different 
magnetic fields and deformations. The results correspond to a dot 
with $N=6$ electrons and $r_s=1.51$.}
\end{figure}


\begin{references}

\bibitem{Sik89} Ch.\ Sikorski and U. Merkt,
Phys.\ Rev.\ Lett.\ {\bf 62}, 2164 (1989).

\bibitem{Dem90} T. Demel, D. Heitmann, P. Grambow, and K. Ploog,
Phys.\ Rev.\ Lett.\ {\bf 64}, 788 (1990).

\bibitem{Brey89} L. Brey, N. F. Johnson, and B. I. Halperin,
Phys.\ Rev.\ B {\bf 40}, 10647 (1993).

\bibitem{Mak90} P. A. Maksym and T. Chakraborty,
Phys.\ Rev.\ Lett.\ {\bf 65}, 108 (1990).

\bibitem{Broi90} D. A. Broido, K. Kempa, and P. Bahshi,
Phys.\ Rev.\ B {\bf 42}, 11400 (1990).

\bibitem{Gud91} V. Gudmundsson and R. R. Gerhardts,
Phys.\ Rev.\ B {\bf 43}, 12098 (1991).

\bibitem{Ser99} Ll.\ Serra {\em et al.},
Phys.\ Rev.\ B {\bf 59}, 15290 (1999).

\bibitem{Pue99} A. Puente, Ll.\ Serra,
Phys.\ Rev.\ Lett.\ {\bf 83}, 3266 (1999).

\bibitem{Ull00} C. A. Ullrich and G. Vignale,
Phys.\ Rev.\ B {\bf 61}, 2729 (2000).

\bibitem{Gudxx} I. Magnusdottir, V. Gudmundsson,
Phys.\ Rev.\ B {\bf 60}, 16591 (1999).

\bibitem{Str94} R. Strenz {\em et al.},
Phys.\ Rev.\ Lett.\ {\bf 73}, 3022 (1994).

\bibitem{Loc96} D. J. Lockwood {\em et al.},
Phys.\ Rev.\ Lett.\ {\bf 77}, 354 (1996).

\bibitem{Sch96} C. Sch\"uller {\em et al.},
Phys.\ Rev.\ B {\bf 54}, 17304 (1996).

\bibitem{Sch98} C. Sch\"uller {\em et al.},
Phys.\ Rev.\ Lett.\ {\bf 80}, 2673 (1998).

\bibitem{Stei99} C. Steinebach, C. Sch\"uller and D. Heitmann,
Phys.\ Rev.\ B {\bf 59}, 10240 (1999).

\bibitem{Stef99} O. Steffens and M. Suhrke,
Phys.\ Rev.\ Lett.\ {\bf 82}, 3891 (1999).

\bibitem{Bar00} M. Barranco {\em et al.}, 
Phys.\ Rev.\ B {\bf 61}, 8289 (2000).

\bibitem{orbital} Ll.\ Serra, A. Puente, E. Lipparini,
Phys.\ Rev.\ B {\bf 60}, 13966 (1999).

\bibitem{inert}
Ll.\ Serra, A. Puente and E. Lipparini, 
submitted for publication (2000).

\bibitem{units}
Unless stated otherwise, we shall use effective atomic units (for 
which $\hbar^2=e^2/\kappa=m=1$, with $\kappa$ and $m$ the dielectric 
constant and electron effective mas, respectively) corresponding to 
GaAs values, i.e., H$^*\approx 12$ meV, $a_0^*\approx 98$ {\AA} and 
$\tau^*\approx 55$ fs, as energy, length and time units.

\bibitem{note} For bulk GaAs the effective gyromagnetic 
factor is $g^*=-0.44$.

\bibitem{Dip94} O. Dippel, P. Schmelcher, and L. S. Cederbaum,
Phys.\ Rev.\ A {\bf 49}, 4415 (1994).


\bibitem{Mad94} A.V. Madhav and T. Chakraborty,
Phys.\ Rev.\ B {\bf 49}, 8163 (1994).

\bibitem{Sakxx} J. J. Sakurai,
{\em Modern Quantum Mechanics}, (Addison-Wesley, Reading 1994),
p.\ 96.

\bibitem{Fer94} M. Ferconi and G. Vignale,
Phys.\ Rev.\ B {\bf 50}, 14722 (1994).

\bibitem{Hei95} O. Heinonen, M. I. Lubin and  M. D. Johnson,
Phys.\ Rev.\ B {\bf 56}, 10373 (1997).

\bibitem{Kos98} M. Koskinen, M. Manninen, S. M. Reimann, Phys.\ Rev.\
Lett.\ {\bf 79} 1389 (1997).

\bibitem{Pi98} M. Pi {\em et al.},
Phys.\ Rev.\ B {\bf 57}, 14783 (1998).

\bibitem{Hi99} K. Hirose, N. S. Wingreen, Phys.\ Rev.\ B {\bf 59}, 
4604 (1999).

\bibitem{Lip99} E. Lipparini {\em et al.},
Phys.\ Rev.\ B {\bf 60}, 8734 (1999).

\bibitem{Reixx} 
S. M. Reimann {\em et al.}, 
Phys.\ Rev.\ Lett.\ {\bf 83}, 3270 (1999).

\end{references}
\end{document}